# On Some Modal Implications of the Dynamic Mode Decomposition: Through the Lens of a Subcritical Prism Wake

Cruz Y. Li[1], Tim K.T. Tse [2*], Gang Hu [3], Lei Zhou [4]

*[1 2 4] Department of Civil and Environmental Engineering, Hong Kong University of Science and Technology, Clear Water Bay, Hong Kong SAR, China*

*[3] School of Civil and Environmental Engineering, Harbin Institute of Technology, Shenzhen, 518055, China*

yliht@connect.ust.hk ORCID 0000-0002-9527-4674

timkttse@ust.hk,

hugang@hit.edu.cn

lzhouau@connect.ust.hk,

E-mail address: timkttse@ust.hk,

Mailing address: Department of Civil and Environmental Engineering, Hong Kong University of Science and Technology, Clear Water Bay, Hong Kong SAR, China

[*] Corresponding Author

## Abstract

The Dynamic Mode Decomposition (DMD) is a Koopman-based algorithm that straightforwardly isolates individual mechanisms from the compound morphology of direct measurement. However, many may be perplexed by the messages the DMD structures carry. This work investigates the modal implications of the DMD/Koopman modes through the prototypical subcritical free-shear flow over a square prism. It selected and analysed the fluid mechanics and phenomenology of the ten most dominant modes. The results showed that the reduced-order description is morphologically accurate and physically insightful. Mode 1 renders the mean-field. Modes 2 depicts the roll-up of the Strouhal vortex. Mode 3 delineates the Bloor-Gerrard vortex resulting from the Kelvin-Helmholtz instability inside shear layers, its superposition onto the Strouhal vortex, and the concurrent flow entrainment. Modes 4, 5, 7, 8, and 9 portray the harmonic excitation. Modes 6 and 10 describe the low-frequency shedding of turbulent separation bubbles (TSBs) and turbulence production, respectively, which contribute to the beating phenomenon in the lift time history and the flapping motion of shear layers. Finally, this work demonstrates the capability of the DMD in providing insights into similar fluid problems. It also serves as an excellent reference for an array of other nonlinear systems.

## Keywords

*Dynamic Mode Decomposition; Reduced-Order Modeling; Subcritical Free-Shear Flow; Prism wake; Koopman Analysis*

# Highlights

- Applied the Dynamic Mode Decomposition (DMD) algorithm to a canonical nonlinear system: the turbulent free-shear flow over prism.
- Underpinned the DMD's ability to dissect the spatiotemporal dynamics of fluid systems by reduced-order modeling for physical insights.
- Revealed the Bérnard-Kármán vortex shedding consists of the superimposing Strouhal vortex and the Bloor-Gerrard vortex.
- Disclosed the low-frequency shedding of turbulent separation bubbles and turbulence production.
- Outlining a methodical reference for future Koopman analysis or DMD applications to nonlinear systems.

# 1. Introduction

In the history of science, analysing nonlinear systems has never been a straightforward task. While nonlinearity remains an intricacy *per se*, many other factors may add to a system's intrinsic complexity. For example, fluid systems, governed by the nonlinear Navier-Stokes equations, are almost always coupled with the issue of high-dimensionality: the stringent requirement for resolving turbulence inevitably pushes the system's degrees of freedom towards infinity. Other factors like the entanglement of dynamics may further obstruct the analysis of such systems. Empirical observations, say on a turbulent flow field, merely depict a morphology resulting from many dynamical activities acting together. Therefore, the disentanglement of morphology, or equivalently, the isolation of the constituent mechanisms, is critical for analysing many nonlinear systems.

Reduced-Order Modeling (ROM) offers an effective analytical route in this aspect. ROMs project the high-dimensional input data onto low-dimensional subspaces, producing mathematically parsimonious and quintessentially genuine representations of the original data. To this end, the data-driven Dynamic Mode Decomposition (DMD) is a relatively new ROM [1]–[3] that determines an optimal sub-spatial representation of the input while considering the dynamical behaviours of that subspace. Unlike its closest cousin, the Proper Orthogonal Decomposition (POD), it reflects both the spatial and temporal correlations of a system, yielding temporally orthogonal modes to describe dominant features. This capacity of the DMD greatly benefits studies on dynamical systems: while the energy of mechanisms may entangle, their periodicities will always remain distinct. Thanks to these features, the DMD has gained notable popularity since its introduction. Kutz *et al.* [3] composed an excellent collage of early DMD applications to nonlinear systems in fluid mechanics, video processing, signal and controls, epidemiology, neuroscience, and finance. On this note, the latest development of the Spectral POD (SPOD) is mathematically equivalent to a statistically optimal ensemble average of the DMD [4], therefore a subordinate variant of the Koopman analysis.

While many have applied the DMD to fluid systems [5], and some have even investigated precisely the configuration of free-shear flow over bluff-body [1]–[3], [6]–[8], their foci often leaned towards the development of the technique, and the configurations served merely as canonically convincing test subjects. The ardour towards the other end of the rope, that is, bridging the mathematical novelty with a broader engineering audience, is perhaps of less capacity. Through this work, we intend to offer such an attempt by investigating some of the modal implications of DMD-generated information, aiming to assist analytical understandings and perhaps encourage more prevalence of the technique in the engineering of nonlinear systems.

Before proceeding further, we cast away any doubts on the link between the modal and physical representations of systems. The DMD belongs to the ROM family, and its root traces back to the Koopman analysis [9]–[15]. The vast collection of previous renderings [5], [6], [9] testifies to the lucid link between modal structures and the physics of fluids. Kutz *et al.* [3] demonstrated the mathematical link between POD-based ROMs like the DMD and the Navier-Stokes equations. On this basis, this work deployed the DMD to perhaps the most prototypical fluid configuration—the turbulent wake of a subcritical free-shear flow over a square prism. We selected the pressure field for exactly the same justifications as our precursory work [14], which ensured the accuracy and feasibility of our DMD endeavour. The input data is simulated by the high-fidelity Large-Eddy Simulations with Near-Wall Resolution (LES-NWR) [16] at $Re = 22{,}000$ and sampled in the statistically stationary state. Subsequently, the DMD algorithm decomposed the raw, entangled measurement into distinct modal structures. We then determined and analysed the ten most dominant DMD modes consulting fluid mechanics and phenomenology, revealing their physical implications and associated streamwise mechanisms. In the flow regime of interest, the spanwise mechanisms are relatively simplistic and well understood [17], therefore spared in this effort.

It is essential to clarify that although the POD is frequently mentioned in comparative terms with the DMD, it is beyond the authors' intention to define one's superiority over the other. Only in particular circumstances may one's performance be compared to that of the other, and such case-dependent conclusions shall not be exaggerated to generalizations.

In compilation, Section 2 describes the methodology including a summary of the DMD algorithm; Section 3 contains preliminary observations on the LES-simulated flow; Section 4 encompasses discussions on the physical implications of the selected DMD modes; Section 5 summarizes the significant findings of this work.

# 2. The Turbulent Free-Shear Flow

## *2.1 Numerical Detail in Brevity*

As the test subject, we numerically generated a fluid domain that immerses an infinite square prism of side length *D*. The configuration effectively prevents axial features and other three-dimensional complications (**Fig**. 1a) [18]–[20]. As the DMD is acutely sensitive to dynamical subtleties, we refrained from prescribing initial perturbations or freestream profiles to avoid artificial turbulence. The prescribed inflow is an incompressible and uniform wind flow ($M<0.3$) with constant fluid properties. The freestream velocity $U_\infty$ and lengthscale *D* yield a global $Re=22,000$. The simulation evolves with a time-step $\Delta t$

$$\Delta t = \frac{\Delta t\ U_\infty}{D} = 8.03\times 10^{-3}, \tag{2.1.1}$$

which meets the Courant-Friedrichs-Lewy criterion $CFL \leq 1.0$ for solving partial differential equations in all evolutionary steps.

While the large eddies are resolved by numerically solving the filtered Navier-Stokes Equations, the subgrid-scale motion is modelled by the Smagorinsky model with the Lilly formulation [21], [22], in which the Smagorinsky constant $C_s=0.1$. Fig. 1a demonstrates the boundary conditions employed in the numerical domain. We selected the spanwise dimension of $4D$ because, for this particular flow regime during the shear layer transition, this dimension suffices in depicting major spanwise dynamics and was adopted by most previous works [17]. However, we remind the readers of the spanwise A, B, and S modes during the 2D-3D transition at $Re=150\text{-}250$ [17], [23], [24], and the large-scale spanwise vortices in the high-*Re* subcritical regime [25], which may require tailored spanwise dimensions.

A finite-volume, segregated, pressure-based solution algorithm was employed for this low-Mach-number incompressible flow. The projection-based method obtains the velocity field from the momentum equation and satisfies the continuity by corrections of the pressure equation. To this end, we chose the Semi-Implicit Method for Pressure Linked Equations-Consistent (SIMPLEC) scheme for the pressure-velocity coupling. We also selected the second-order scheme for the pressure interpolation, the second-order scheme for the viscous terms of spatial discretization. The simulation was initially performed with the third-order Quadratic Upstream Interpolation for Convective Kinematics (QUICK) scheme for the discretization of convection term, in which the upstream correction parameter $\theta$ was set to $\theta \leq 0.125$. We then validated the QUICK results with the second-order bounded central-differencing scheme and observed minimal statistical differences.

As this work is a continuing effort of our precursory work on the feasibility and accuracy of the DMD decomposition [14], we inherited the LES-NWR simulation therein. To avoid redundancy, readers are directed to Li *et al.* [14] for full numerical details.

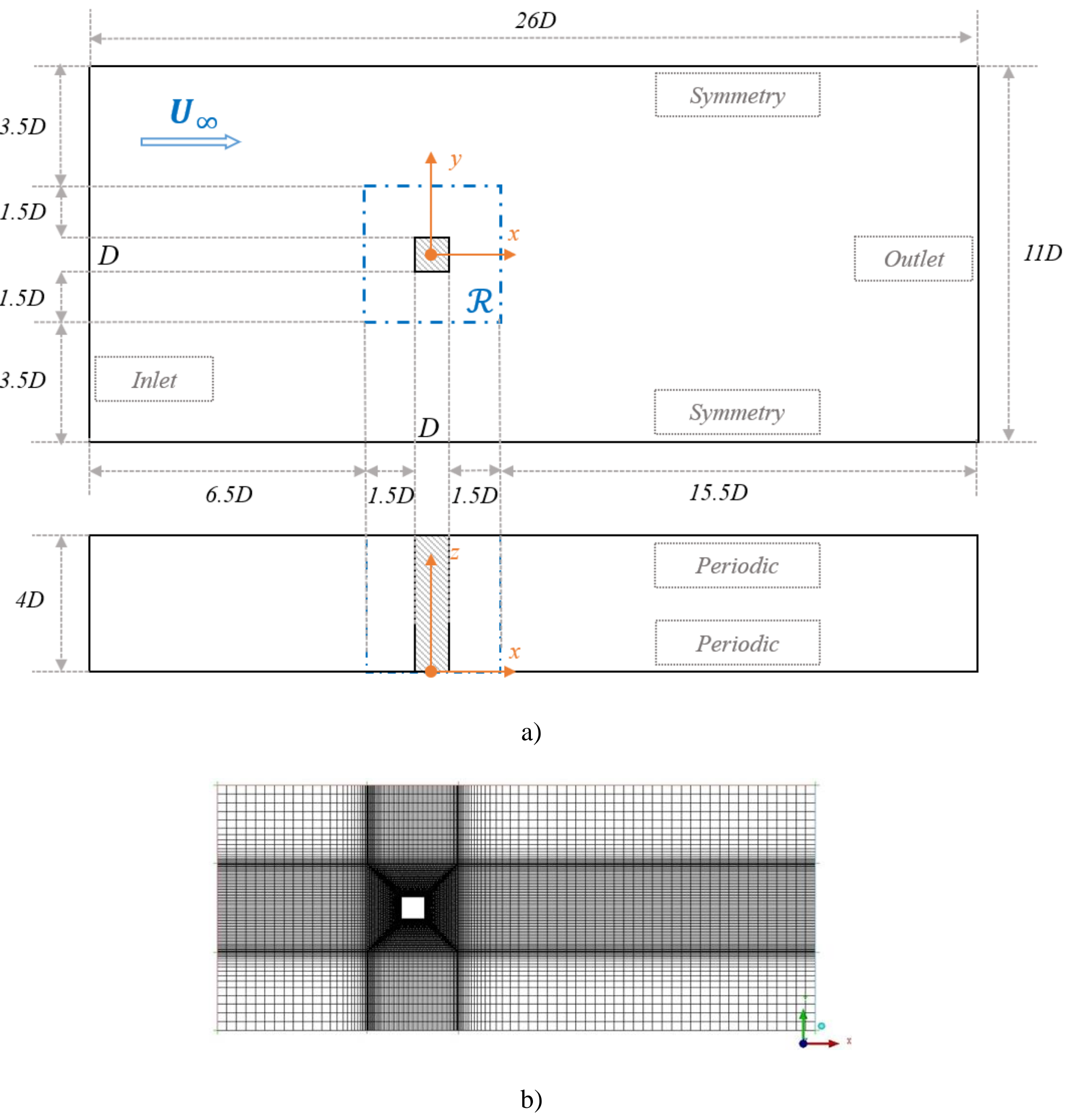

**Fig. 1 a)** The numerical fluid domain; **b)** The hexahedral grid.

## *2.2 Grid Validation*

The fluid domain was discretized into a strictly-structured hexahedral, *4.5* million grid to minimize numerical error. We also imposed a refinement region $R$ (Fig. 1b), enclosing the prism, incoming flow, separation zone, and the near wake to ensure the near-wall resolution. This work estimated the thickness of the viscous sublayer to determine the height of the wall-adjacent cells. The estimation relied on an analogy of the laminar boundary layer on a flat plate, as pointed out by [26]:

$$\delta_{BL,\,99\%} \approx 3.5\sqrt{\frac{2\,\nu\,l_x}{U_\infty}} \approx \frac{5.0\,l_x}{\sqrt{Re_x}} = 0.024D \qquad (2.2.1)$$

where $l_x$ denotes the characteristic length of the boundary layer, which is taken as $0.5D$ after Cao *et al.* [27]. We set a fine resolution of $^{1}/_{4000}\,D$ to keep the $y^+$ strictly under unity and a grading ratio of $1.08$ to resolve the viscous sublayer by as many as $32$ layers. Finally, the grid construction followed Menter [28], the COST 732 Code [29], and the AIJ Guideline [30] for resolution in the $x$- and $z$-directions.

This work generated three grid systems with $2.0$, $4.5$, and $10.1$ million cells to establish grid independence, which proved the selected $4.5$ million grid suffices. However, this criterion was initially developed for the Reynolds-Averaged Navier-Stokes (RANS) implementations. Though often retained as a CFD convention, its applicability to an LES-NWR grid is subject to debate. Therefore, in addition to the validation presented in our previous work [14], we offer additional evidence to consolidate the grid resolution. Since the LES-NWR solves the filtered velocity field by the numerical Navier-Stokes equations, the resolution of the grid directly determines that of the filtering, hence the accuracy of the simulation. To this end, a grid-dependent filter in the physical space $\Delta = V^{1/3}$ was adopted, where $V$ denotes the cell volume.

As Pope pointed out in his masterwork on turbulent flows [16], a proper LES-NWR shall have its filter placed in the inertial subrange to fully resolve the energy-containing range. Otherwise, the large, energy-containing eddies are inappropriately modelled by the subgrid-scale model in the form of Very-Large-Eddy-Simulations (VLES). Pope [16] also defined a demarcation between the energy-containing range and the inertial subrange -- 80% resolution for the total turbulence kinetic energy $\langle k \rangle$.

In what follows, the resolved portion of $\langle k \rangle$ is denoted by $k_r$, the subgrid portion by $k_{sgs}$, and the numerical portion by $k_{num}$. The resolved spectrum, $E$, is expressed by

$$E \equiv G(x)\,E_{\langle k \rangle} \approx \frac{k_r}{\langle k \rangle}, \qquad (2.2.2)$$

$$\langle k\rangle=k_r+k_{sgs}+k_{num}, \tag{2.2.3}$$

$$k_r=\frac{1}{2}\left(\overline{u'^2}+\overline{v'^2}+\overline{w'^2}\right) \tag{2.2.4}$$

$$k_{sgs}={v_{sgs}}^2/{l_s}^2 \tag{2.2.5}$$

$$l_s=min(\kappa d,C_s\Delta) \tag{2.2.6}$$

where $G(x)$ denotes the filter function in three-dimensional space; $\overline{u'^2}$, $\overline{v'^2}$, and $\overline{w'^2}$ denote the variance of the fluctuating velocities $u'$, $v'$, and $w'$, respectively. $k_{num}$ is a pseudo-energy term that accounts for discretization error and numerical residual. Celik *et al.* [31] pointed out that $k_{num}$ is sufficiently small for an LES-NWR with an overall second-order discretization. As introduced before, our discretization is at least second-order, producing minimal numerical dissipation. As the Courant-Friedrichs-Lewy (*CFL*) condition is strictly maintained, the numerical dispersion is also insignificant. Furthermore, our convergence criteria are as stringent as $1\times10^{-6}$ for both the continuity and momentum equations. Therefore, $k_{num}$ is deemed negligible.

Fig. 2 presents the mid-plane time-averaged resolved spectrum in the statistically stationary state. Evidently, the entire fluid domain satisfies the requirement of *80%*. The global minimum of *82.6%* signifies that even at locations with extreme local Reynolds numbers, *i.e.,* the wall jets immediately after separation and the near-wall viscous region, the filter rests in the inertial subrange and resolves the entire energy-containing range. The stringent quality assessment ensures the reliability of the LES-simulated flow field. To avoid repetition, we direct readers to [14] for other conventional validation procedures.

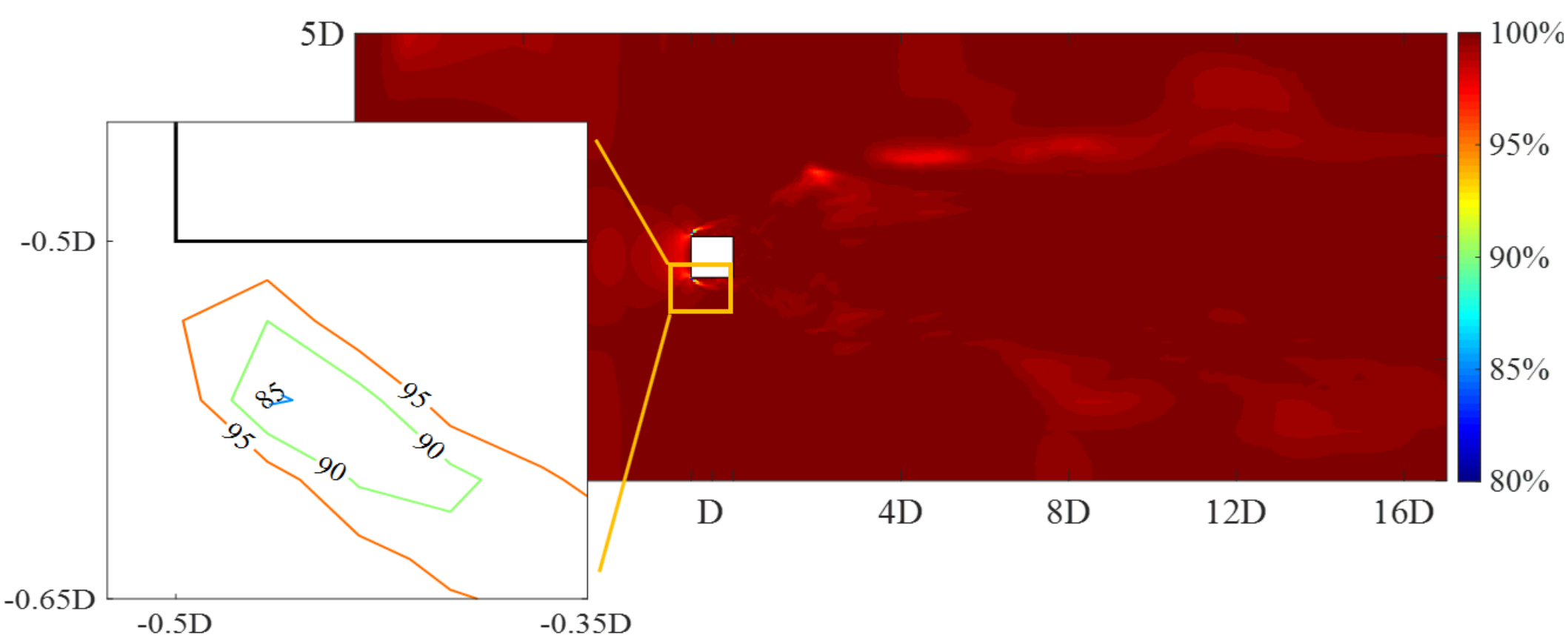

**Figure 2** Time-averaged resolve spectrum of turbulent kinetic energy in statistically stationary state.

## *2.3 Statistical Stationarity*

During the simulation, we selected three points that characteristically represent the stagnation (Point 1), shear layer dynamics (Point 2), and prism base (Point 3) to monitor the local statistics (Fig. 3a). At all three points, the normalized mean velocity magnitude reached approximate asymptotes after *60,000* time-steps, embodying the definition of statistical stationarity (Figs. 3b-d). Likewise, the time history of global lift coefficient (Fig. 3e) shows that after *60,000* time-steps, transient processes in the initialization stage diminished, while the mean and root-mean-square (r.m.s.) values reached their respective asymptotes. The *60,000* mark also corresponds to *20* flow-through-times, which further buttresses the statistical stationarity according to Iousef *et al.* [32].

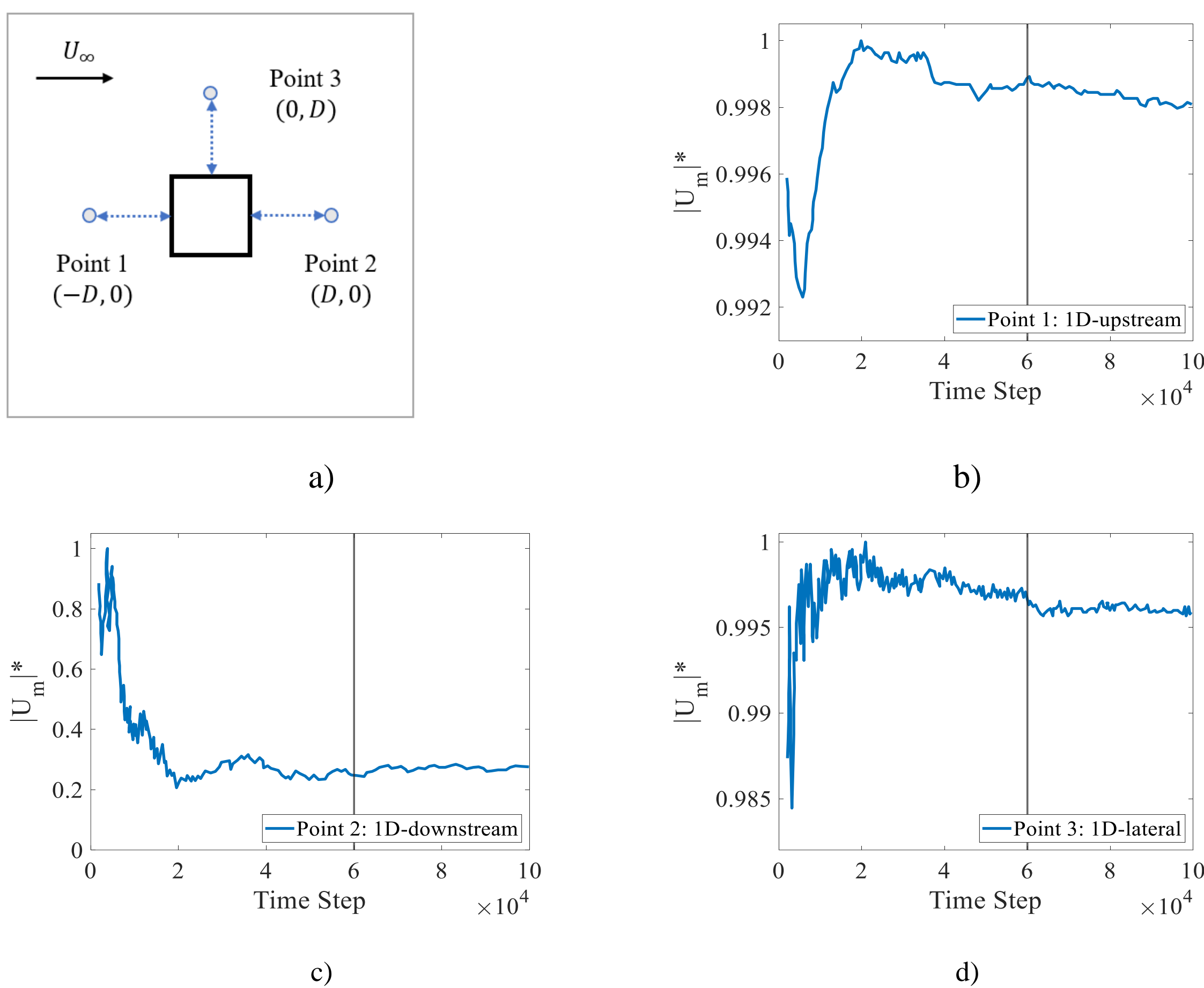

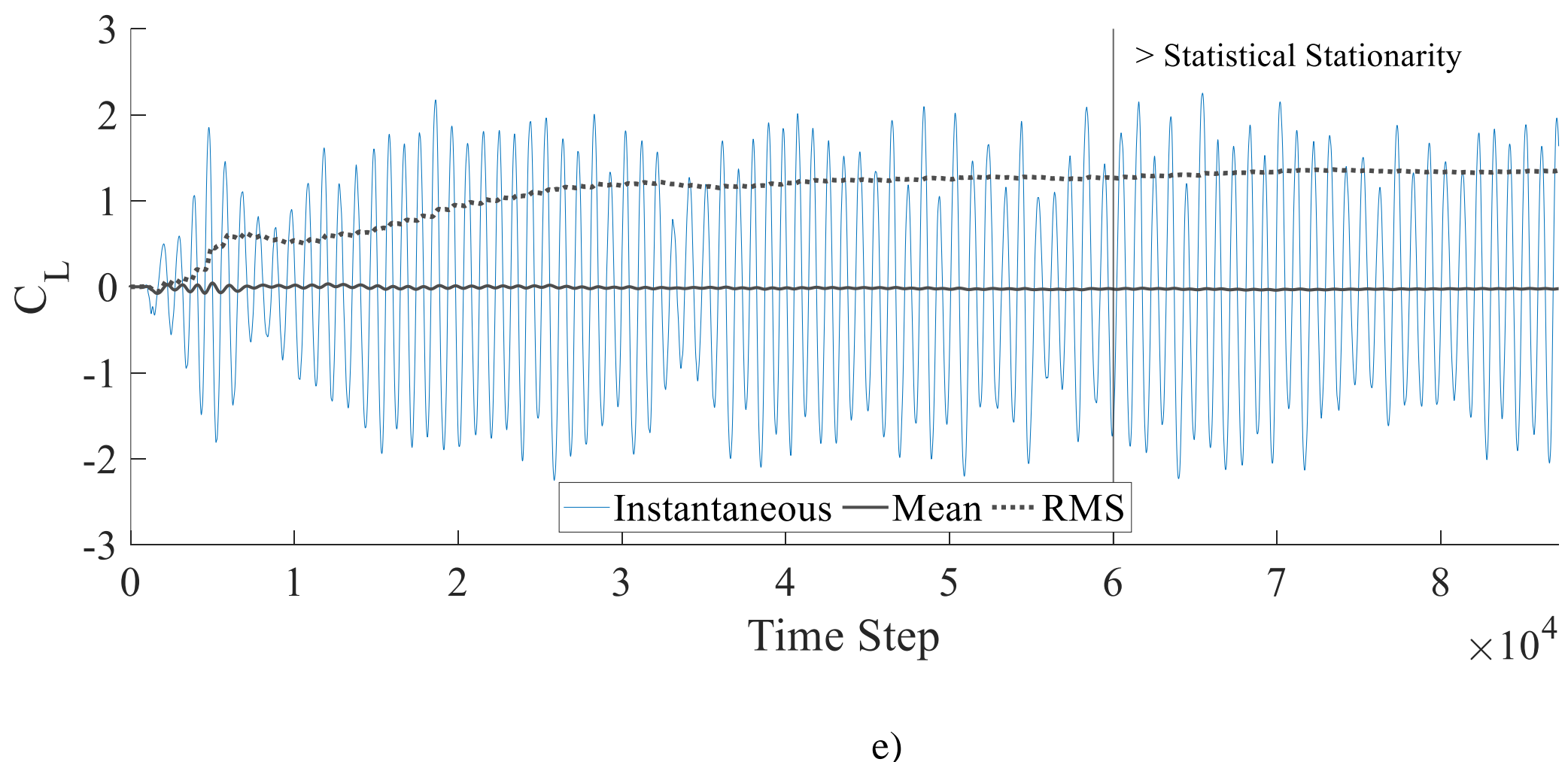

e)

Figure 3 a) Locations of three monitoring points; Normalized mean velocity magnitude of b) Point 1, c) Point 2, d) Point 3; e) Time history of instantaneous, mean, and r.m.s. lift coefficient.

Since the LES captures unsteady pressure components, we expect fluctuations in force coefficients (Fig. 3e). More importantly, a beating phenomenon is clearly observed in the time history. According to Muld *et al.* [33] and Cesur *et al.* [34], the beating phenomenon is associated with a low-frequency, oscillatory behaviour of the wake and shear layers, which persists throughout the statistically stationary state. We will further discuss this notion in subsequent discussions.

Based on instantaneous lift, we performed a power spectral density (PSD) analysis (Fig. 4). Evidently, a prominent peak at $St_{vs} = f_{vs}D/U_\infty = 0.126$ captures the the Bérnard-Kármán vortex shedding. Additionally, some less prominent peaks are visible between $St = 0.2\text{-}0.4$. These peaks are associated with higher harmonics of a fundamental frequency. Another relatively prominent peak appears at approximately $1/2\ f_{vs}$. This low-frequency activity is of notable significance and will be further dissected using the DMD in Section 4.2.4.

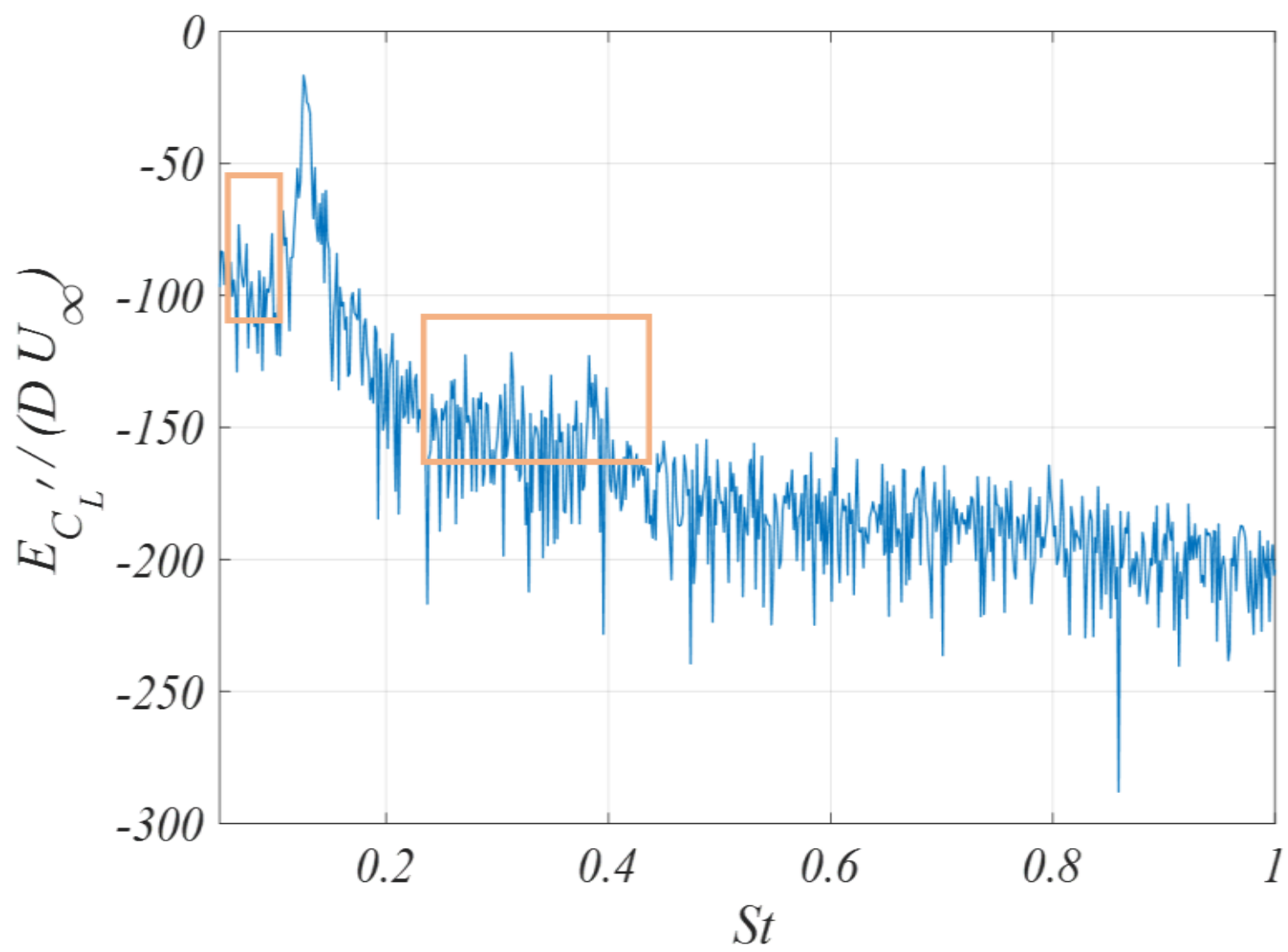

**Figure 4** Power spectral density analysis of the instantaneous lift coefficient

# 3. Dynamic Mode Decomposition

The standard formulation of the Dynamic Mode Decomposition was originally introduced by Schmid [1] and Rowley *et al.* [9]. The DMD algorithm consists of four conceptually quintessential components: data pre-processing, spatiotemporal mapping, modal characterization, and dominant mode selection. This section introduces the four components with brevity. A detailed formulation is presented in Appendix A and Li *et al.* [14].

## *3.1 Data Curation*

The first step is arranging the input data into the prescribed format. The DMD is a data-driven, physics-uninformed algorithm, so its input data can essentially be of any nature. Nevertheless, the input data for the vanilla algorithm shall be:

1) instantaneous for meaningful results,
2) in discrete-time with a constant sampling frequency,
3) captured on a fixed domain, as preliminary tests found alterations in the data domain led to numerical degenerations.

For unqualified data, readers are directed to [1], [3] for non-uniform sampling and [35] for a non-uniform domain.

Upon acquisition, the input data may need pre-processing like interpolation and noise elimination. While errors may exist, noise is typically non-existent for CFD data. However, for empirical data, noise elimination is crucial, as the DMD is extremely sensitive to frequency content and is sometimes unable to distinguish noise from the high-frequency dynamics. Over-contamination by noise may lead to the total loss of decomposition integrity, especially for highly turbulent flows. Readers may refer to [36], [37] for noise elimination techniques for the DMD.

The final step of data pre-processing is to arrange the input data into two snapshot sequences in matrix form. In this work, the sequences contain snapshots of the instantaneous pressure field. Each snapshot $\boldsymbol{x}_i$ consists of nodal pressure values. The snapshots $\boldsymbol{x}_i$ are sampled in the uniform LES sampling time $\Delta t$, which resolves each vortex shedding cycle by *991* times steps. The two snapshot sequences are arranged as:

$$\boldsymbol{X}_1=\{\boldsymbol{x}_1,\boldsymbol{x}_2,\boldsymbol{x}_3,\ldots,\boldsymbol{x}_{m-1}\}, \tag{3.1.1}$$

$$\boldsymbol{X}_2=\{\boldsymbol{x}_2,\boldsymbol{x}_3,\boldsymbol{x}_4,\ldots,\boldsymbol{x}_{m}\}, \tag{3.1.2}$$

where $\boldsymbol{x}_i \in \mathbb{C}^N$; $M$ denotes the temporal length of the snapshot sequence, and $N$ denotes the spatial dimension of each snapshot.

## *3.2 Spatiotemporal Mapping*

As the centrepiece of the technique, the DMD imposes a finite-dimensional matrix to connect the phase-shifted snapshot sequences and approximate the infinite-dimensional linear Koopman operator. Therefore, the DMD is one of many algorithms that approximate the Koopman eigen tuples. So, a DMD mode approximates a Koopman mode. The mapping matrix $\boldsymbol{A}$ contains all the spatiotemporal information needed to evolve the first sequence to the second sequence, as shown below:

$$\boldsymbol{X}_2= \boldsymbol{A}\boldsymbol{X}_1. \tag{3.2.1}$$

Expectedly, when the input sequences contain only linear dynamics, the mapping matrix $\boldsymbol{A}$ is exact. If there is any nonlinearity in the input data, $\boldsymbol{A}$ becomes a globally linearized best-fit approximation of the system dynamics.

The remaining effort targets the acquisition of $\boldsymbol{A}$. The vanilla DMD acquires $\boldsymbol{A}$ by either the companion matrix expression [1] or the similarity matrix expression [2]. In practice, the former can be analytically intractable when the state dimension is large (*e.g.,* a high-dimensional flow field) [3]. The latter relies on a similar matrix to project the Koopman operator onto a POD subspace, which is more robust for high dimensional systems [38]. This work formulates the vanilla DMD algorithm using the similarity expression, and the mathematical details are presented in Appendix A2.

### *3.3 Modal Characterization*

The acquisition of the mapping matrix $\boldsymbol{A}$ signals one's full possession of the spatiotemporal information, or its best global linear approximate, of the input data. In this specific implementation, $\boldsymbol{A}$ implicitly representatives the Navier-Stokes Equations. The final stage of the DMD algorithm is to extract modal characteristics of $\boldsymbol{A}$ by an eigendecomposition, as shown below:

$$\widetilde{\boldsymbol{A}}\boldsymbol{W}=\boldsymbol{W}\boldsymbol{\Lambda}, \tag{3.3.1}$$

where $\boldsymbol{W}$ contains the eigenvectors $\boldsymbol{w}_j$, and $\boldsymbol{\Lambda}$ contains the corresponding eigenvalues $\boldsymbol{\lambda}_j$.

The eigen tuples are key to obtain the DMD mode $\boldsymbol{\phi}_j$, continuous-time frequency $\omega_j$, and growth/decay rate $g_j$ respectively. Mathematical details to retrieve the DMD parameters are presented in Appendix A3.

### *3.4 Dominant Mode Selection*

We employed the $\alpha$-criterion [3] to select the most dominant DMD modes and the $I$-criterion [38] for validation. In this process, we ranked all DMD modes by their modal contributions after evaluating their spatial (*i.e.*, energy) and temporal behaviours. Therefore, DMD modes with the highest modal contributions describe the most dominant flow mechanisms in the overall flow dynamics.

### *3.4.1 α-criterion*

The $\alpha$-criterion relies on the coefficients of weight, or the $\alpha$ amplitude, to determine a DMD mode's modal contribution. The $\alpha$ amplitude is determined by projecting DMD modes onto the first snapshot, with which the exponential term at $t = 0$ vanishes, as shown below:

$$\boldsymbol{\alpha}=\boldsymbol{\Phi}^{\dagger}\boldsymbol{x}_1 \tag{3.4.1.1}$$

$$\boldsymbol{x}(t) = \sum_{i=1}^{N} \phi_i e^{\omega_i t}\alpha_i = \boldsymbol{\Phi} e^{\boldsymbol{\Omega} t}\boldsymbol{\alpha} \tag{3.4.1.2}$$

where $\boldsymbol{\Phi}^{\dagger}$ denotes Moore-Penrose pseudoinverse,

Thereafter, the temporal quantity time evolution $T_j(t)$ of DMD mode $\boldsymbol{\phi_j}$ can be obtained from:

$$T_j(t)= e^{\omega_j t} \tag{3.4.1.3}$$

### *3.4.2 I-criterion*

The $I$-criterion is an improved mode selection criterion that evaluates the time dynamics of the DMD modes in addition to their α amplitude. To this end, a Vandermonde matrix $\boldsymbol{V_{and}}$ describing the time dynamics of DMD modes is introduced:

$$\boldsymbol{X}_1=\boldsymbol{\Phi D_{\alpha}} V_{and}= \left[\boldsymbol{\phi}_1, \boldsymbol{\phi}_2, \ldots, \boldsymbol{\phi}_r\right] \begin{bmatrix} \alpha_1 & & & \\ & \alpha_2 & & \\ & & \ddots & \\ & & & \alpha_r \end{bmatrix} \begin{bmatrix} 1 & \lambda_1 & \cdots & \lambda_1^{i-1} \\ 1 & \lambda_2 & \cdots & \lambda_2^{i-1} \\ 1 & \vdots & \ddots & \vdots \\ 1 & \lambda_r & \cdots & \lambda_r^{i-1} \end{bmatrix} \tag{3.4.2.1}$$

where $i$ denotes time instant in the input signal.

One derives the $I$-amplitude $I_j$ by:

$$I_j=\sum_{i=1}^{N} \left|\alpha_j \lambda_j^{i-1}\right| \left\|\boldsymbol{\phi}_j\right\|_F^2 t* \tag{3.4.2.2}$$

where the brace $\| \; \|_F$ denotes Frobenius normalization.

# 4. Results and Discussions

## *4.1 Dominant DMD Mode Selection*

This section presents the selection of the ten most dominant DMD Modes using the α-criterion, which the *I*-criterion also validates. By the vanilla DMD formulation, a full-order DMD model yields *N* modes corresponding to the temporal dimension. We emphasize that not every generated *mode* is a DMD *Mode*. For an oscillatory DMD Mode, a conjugate mode pair is generated so that each component pertains to the real or imaginary conjugate of an eigenfrequency. Consequently, it consists of two modes with either the positive or negative counterpart of a physical frequency that share identical growth/decay rate and modal contribution. For a non-oscillatory DMD Mode, no imaginary conjugate exits. In subsequent discussions, we denote a conjugate of an oscillatory DMD Mode by the small letter *mode*, and the non-oscillatory or oscillatory DMD Mode *per se* by the capitalized *Mode*.

The ten most dominant DMD Modes are ranked by their $\alpha$ amplitude (Fig. 5a). Clearly, Mode 1 is a non-oscillatory standalone mode at *0 Hz*, whereas Modes 2-10 are all oscillatory modes in conjugate pairs. The conjugates are symmetric about the *0-Hz* abscissa, but only the positive conjugate holds physical meaning (Fig. 5b).

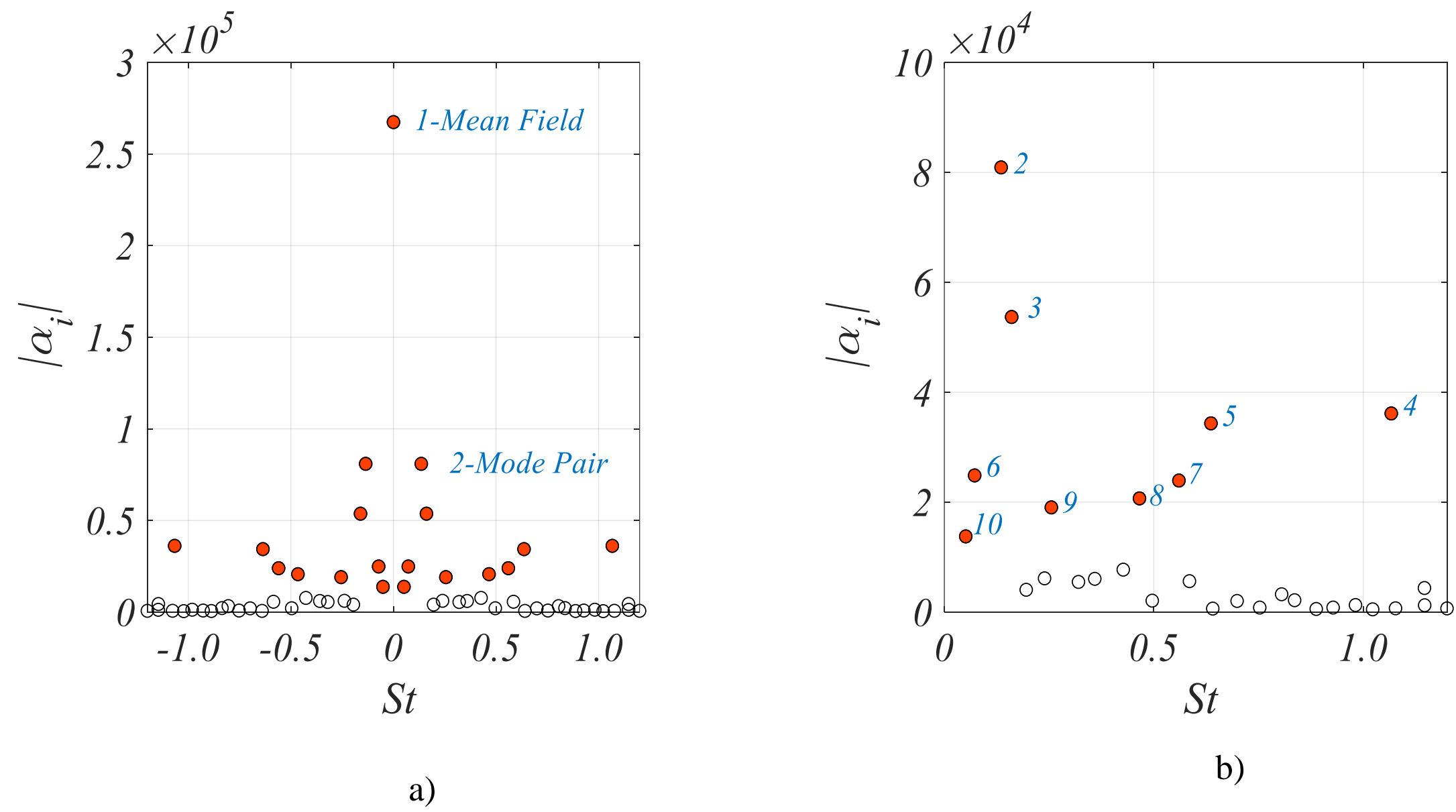

**Figure 5 a)** Modal amplitude by the α-criterion; **b)** the 10 dominant modes.

An essential step to mode selection is the assessment of mode stability. For this purpose, we present the DMD spectrum in Fig. 6a. The real-imaginary plane displays the *Region of*

*Convergence* (ROC) for the dynamical system. In essence, the finding of DMD modes is almost equivalent to locating the poles of a system from a purely signal processing perspective. The fact that all DMD modes lie closely along the unit circle with very few exceptions indicates system stability. The proximity to the unit circle is also an indicator of mode stability. Thus, Modes 1-10 are satisfactorily stable, and the overall DMD sampling sufficiently converged. However, close inspection revealed that Mode 4 is subjected to inferior stability relative to its peers (Fig. 6b). Although it is imprudent to disregard Mode 4 altogether, this observation will play a part in the subsequent analysis.

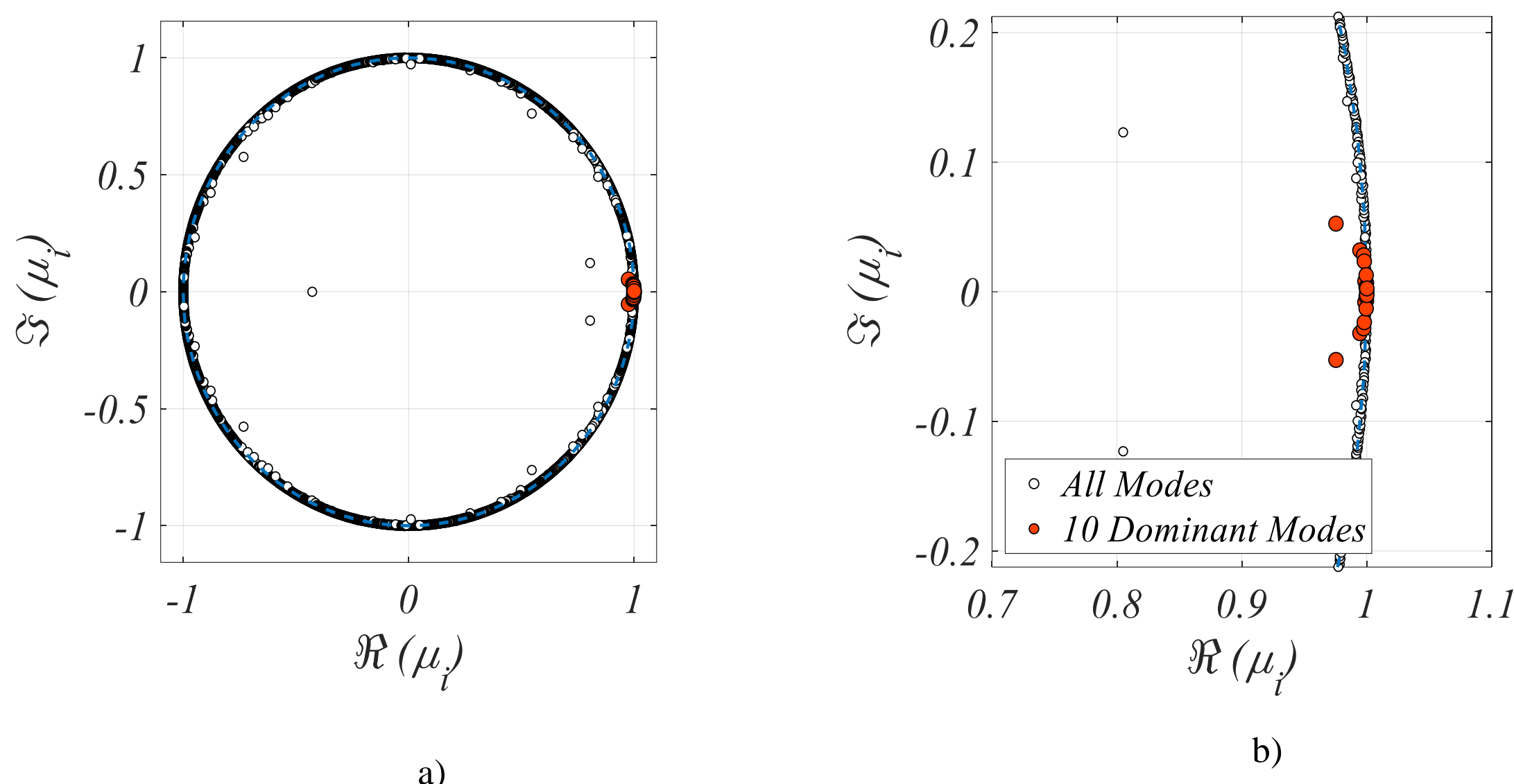

**Figure 6** a) DMD spectrum on the $\Re$-$\Im$ plane; b) Zoomed-in spectrum.

The same selection process is replicated for the *I*-criterion, which yields the same dominant Modes to validate the $\alpha$-criterion. The similarity means both criteria agree on spatiotemporal dominance of the selected Modes. For concision, the *I*-criterion selection process is omitted from this presentation, but we present relevant information in Table 1.

At this point, we emphasized that the number 'ten' for the mode selection is not at all arbitrary. Based on the modal contribution of the ranked DMD Modes (Table 1), this work selects *2.00%* as the cut-off threshold. Both selection criteria show that the first ten Modes meet this requirement. We also clarified that *2.00%* is not a definitive demarcation that decides modal dominance. Instead, it is merely a cut-off imposed to preserve the concision of presentation. Users may adjust this threshold according to analytical needs.

**Table 1** Summary of dominant Modes 1-10.

| Rank | $\alpha_j$ [$10^3$] | $\alpha_j$ contribution [%] | $I_j$ [$10^3$] | $I_j$ contribution [%] | Growth/decay rate [$10^{-3}$] | *St* |
|---|---|---|---|---|---|---|
| 1 | 267 | 38.8 | 4.05 | 39.1 | -4.3 | 0 |
| 2 | 80.9 | 11.8 | 1.22 | 11.8 | -17.5 | 0.126 |
| 3 | 53.7 | 7.80 | 0.811 | 7.83 | -140.0 | 0.131 |
| 4 | 36.1 | 5.25 | 0.534 | 5.16 | -2347.0 | 1.067 |
| 5 | 34.3 | 4.99 | 0.517 | 4.99 | -490.8 | 0.636 |
| 6 | 24.9 | 3.61 | 0.376 | 3.63 | -45.8 | 0.072 |
| 7 | 23.9 | 3.48 | 0.361 | 3.49 | -202.2 | 0.560 |
| 8 | 20.7 | 3.00 | 0.312 | 3.02 | -165.0 | 0.466 |
| 9 | 19.0 | 2.77 | 0.288 | 2.78 | -34.5 | 0.256 |
| 10 | 13.8 | 2.00 | 0.208 | 2.01 | 10.8 | 0.051 |
| Sum | - | 83.5 | - | 83.8 | - | - |

In terms of the cumulative modal contribution, the ten dominant DMD Modes contribute 83.5% and 83.8% of the total modal amplitude based on the $\alpha$- and $I$-criteria, respectively (Fig. 7). This is to say, the flow mechanisms described by the ten Modes are the primary contributors to the spatiotemporal dynamics of the flow field and constitute most of the flow phenomenology. Meanwhile, the exponential growth of the cumulative modal contribution plateaus as the mode number increases, indicating the lessening dominance of higher-order Modes.

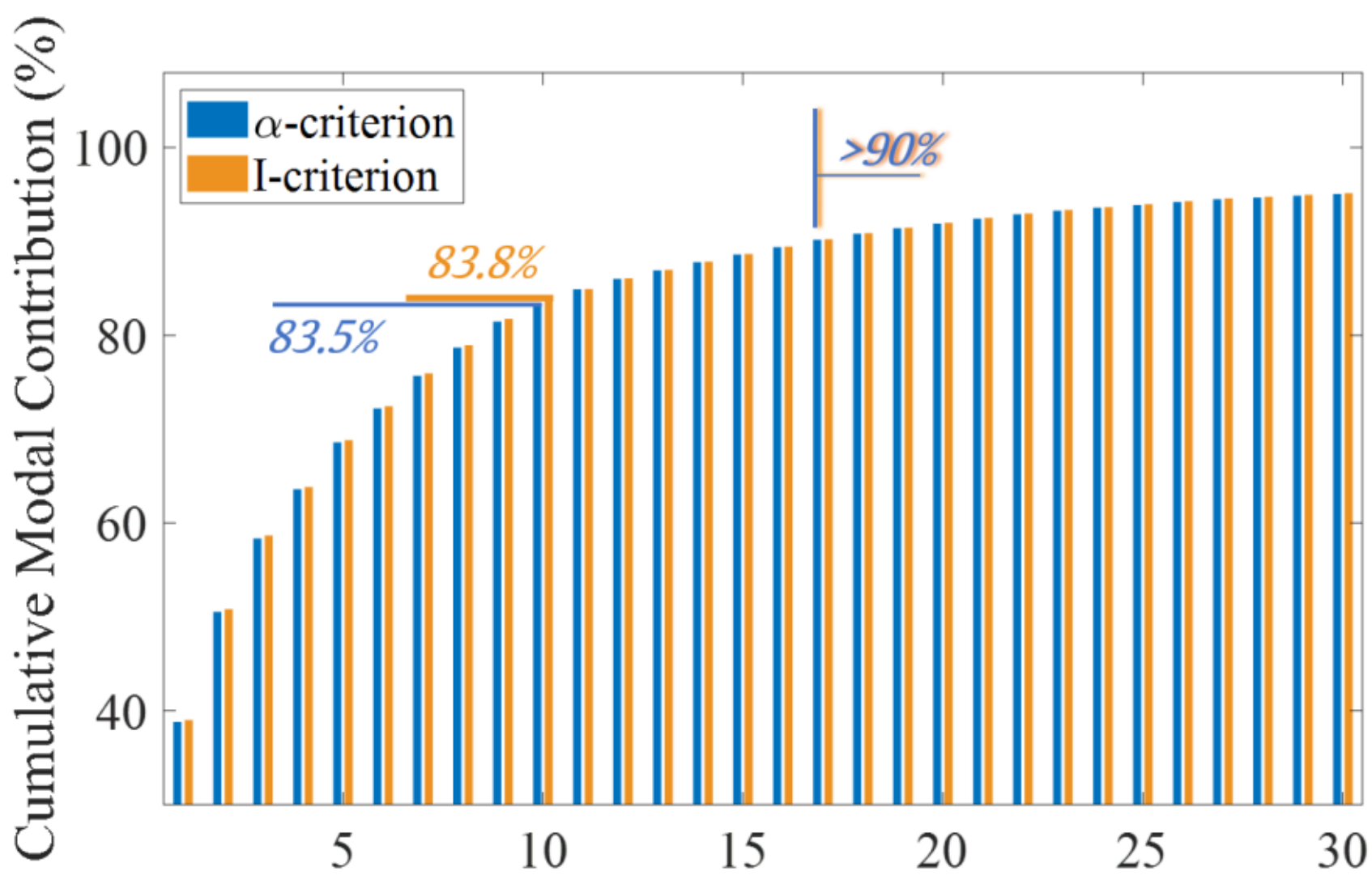

**Figure 7** Cumulative modal contribution of dominant Modes.

## *4.2 Spatiotemporal Analysis and Physical Implications*

One of the most attractive features of the DMD is its ability to linearly and globally approximate complex nonlinear dynamics based on space and time, making previously unobtainable spatiotemporal dynamics accessible with visualisations. In this section, we extract and analyse the physical implications of the ten dominant Modes. The success consolidates the DMD's adequacy for mechanism revelation in complex fluid systems.

### *4.2.1 Mean-field – Mode 1*

In the most straightforward case, Mode 1 describes the mean pressure field. Mode 1 has 40% of modal contribution—the most among all Modes—embodying the reservoir of kinetic energy in a turbulent flow. The process of the Richardson-Kolmogorov energy cascade begins with the extraction of the kinetic energy by the largest eddies, through the processes of turbulence production, from the mean-field. Temporally, the frequency of Mode 1 is $St_1 = 0$, indicating the Mode pertains to no periodicity. Spatially, the mode shape (Fig. 8) displays the time-averaged mean pressure field in the statistically stationary state, as widely reported by previous DMD works [3], [33], [39], [40].

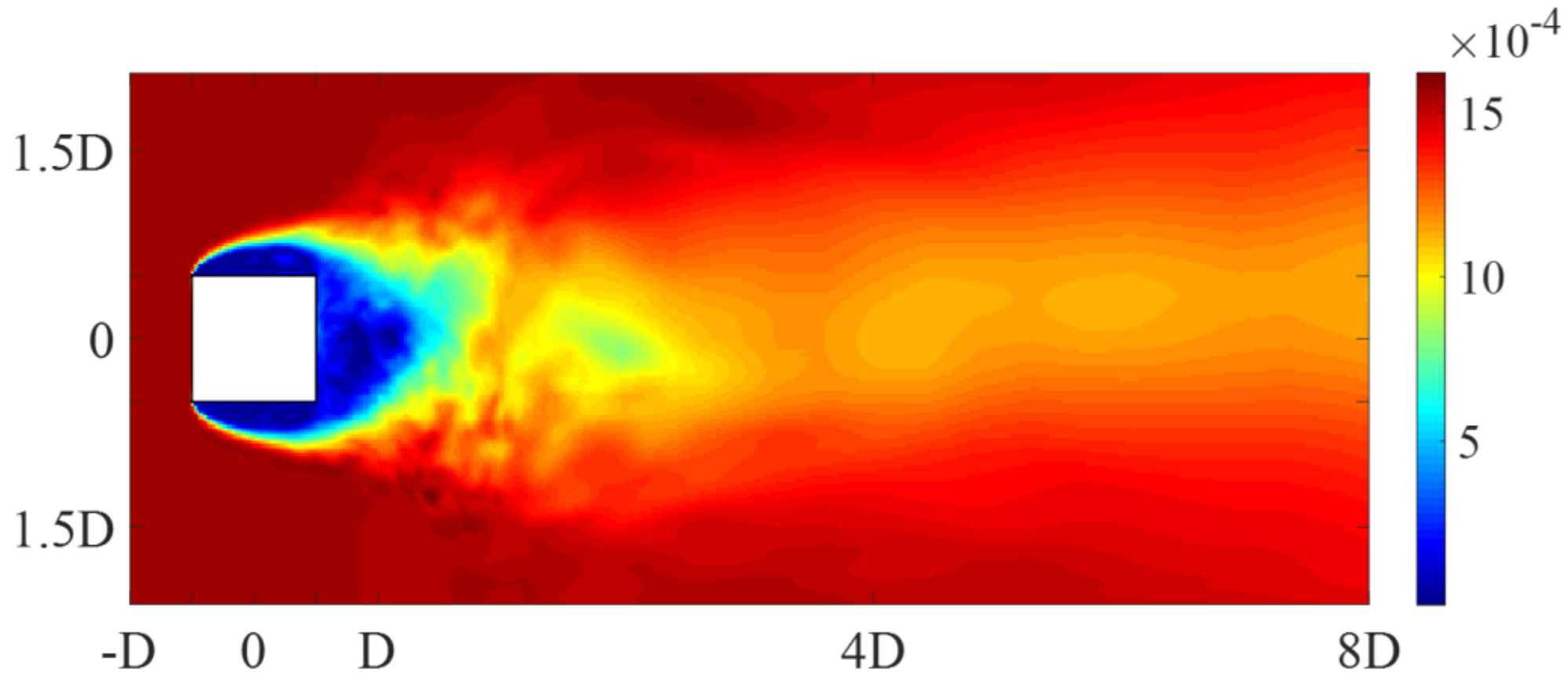

**Figure 8 a)** Mode shape of Mode 1 ($St_1 = 0$).

Perhaps the only unexpected characteristic of Mode 1 is its temporal behaviour. A mean-field DMD Mode typically has a zero-growth/decay. However, Mode 1 has a decay rate of $-4.3\times10^{-3}$. Luo and Kareem [35] also observed the non-zero growth/decay rate in their DMD implementation. They remarked that the mean-field is monotonically increasing where a non-zero growth/decay rate is expected to represent an energy injection that balances the dissipation. We only partially agree with this explanation. The sampling of a turbulent flow, even in the statistical stationary state, is never instantaneously steady. Therefore, a stringent zero growth/growth rate is only possible when sampling time reaches infinity, such that the infinitesimal temporal variations in the statistics complete converge. In practice, the mathematical stringency translates to an infinitely large sample size, which is unpragmatic. Therefore, a non-zero but low growth/decay rate is the anticipated artefact of any pragmatic implementation of the DMD. To this end, $-4.3\times10^{-3}$ is sufficiently low and attests to the sampling convergence.

Readers are also reminded that mean-subtraction is another way to cope with non-zero growth/decay rate. Chen et al. [41] showed that mean-subtracted DMD modes are mathematically equivalent to Discrete Fourier Transform (DFT) modes. DFT modes often provide better approximations of the Koopman modes [4], but mean-subtraction is not mandatory for the Koopman analysis.

### 4.2.2 Bérnard-Kármán Vortex Shedding – Modes 2 and 3

We analyse Modes 2 and 3 in conjunction because they are the most dominant oscillatory Modes. Modes 2 and 3 are associated with the Bérnard-Kármán vortex shedding (BKVS) phenomenon. Temporally, the physical frequencies of Modes 2 and 3 are $St_2 = 0.126$ and $St_3 = 0.131$, respectively, which closely match $St_{vs}$ (Fig. 4). On the other hand, the BKVS is the well-known predominant mechanism for this canonical free-shear flow configuration. The dominance of Modes 2 and 3 in the system's dynamics echoes that of the BKVS in the excitation mechanisms.

However, before leaping to conclusion faithfully, one question must be answered: why does the BKVS decompose into two DMD Modes with distinct frequencies? To answer this question, we resorted to fluid mechanics and flow phenomenology. The vorticity generated by the viscous effect inside the recirculation zone and that developed within the shear layers result in two distinct kinds of vortices [42]. The BKVS takes place only as the two kinds of vortices interact. Perhaps the first kind draws more familiarity: vorticity arises from no-slip walls, convects downstream into the wake, and triggers the roll-up of shear layers into large, counter-rotating vortices. Wu *et al*. [43] refer to this kind of vortex as the *Strouhal vortex*. Hussain [44] refers to it as the *rolls*.

On the other hand, another kind of small-scale vortex found in the shear layers is named the *Bloor-Gerrard (BG) vortex* after fluid mechanists M. S. Bloor [45] and J. H. Gerrard [46]. The Bloor-Gerrard instability is established based on the Gerrard theory [46], [47] of bluff body wakes and originates from the Kelvin-Helmholtz (KH) instability [48] in the shear layers. Hussain [44] also refers to them as the *ribs* Accordingly, the BG vortices are belted by shear layers into the prism base. In the Kármán Street, the two kinds of vortices *superimpose* but do not *superpose* onto one another, so they are independently detected by experiments [43], [49]. Therefore, the detection of two DMD Modes also makes intuitive sense [50].

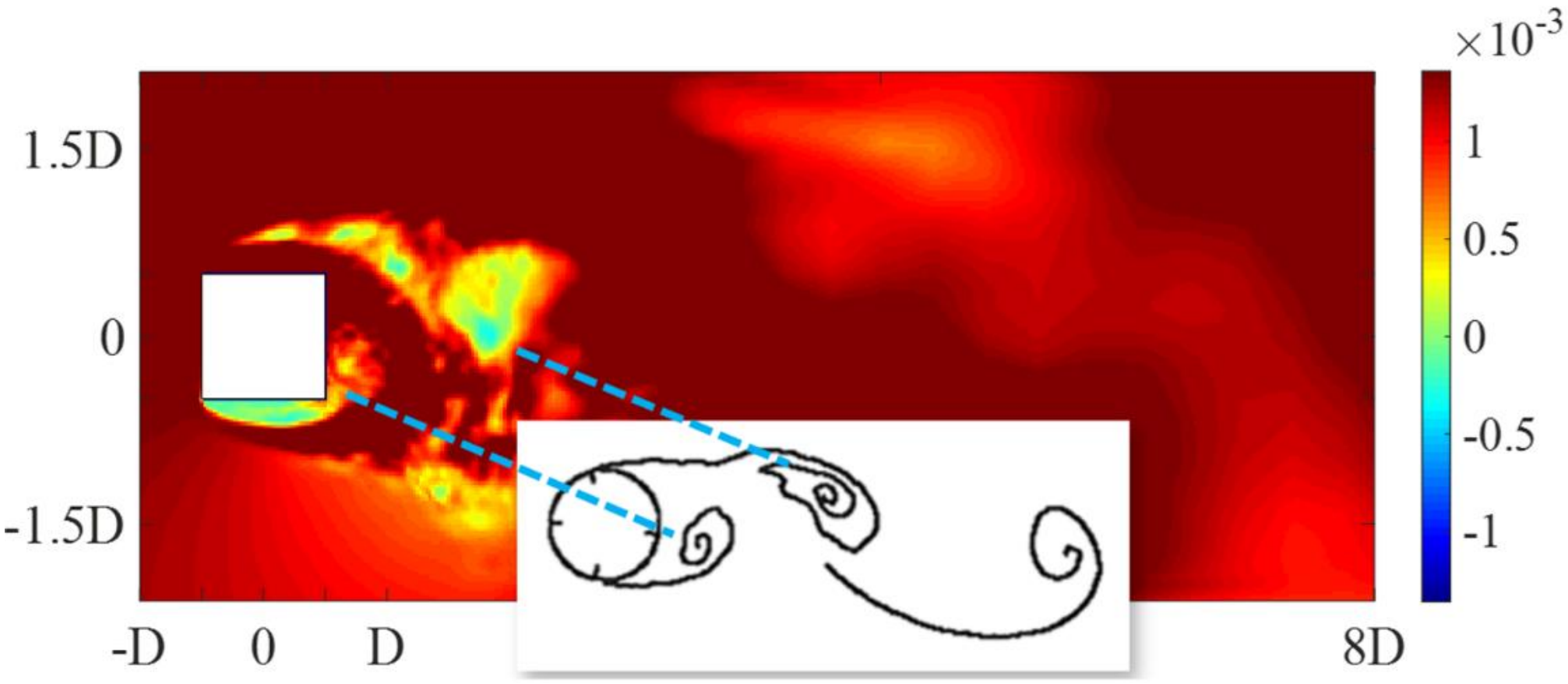

**Figure 9** Mode shape of Mode 2 compared to the schematic illuration of the formation of a Strouhal vortex from Sarkpaya [51] ($St_2 = 0.126$)

The physical implications of Modes 2 and 3 become lucid. The oscillatory Mode 2 describes the formation of the Strouhal vortex, as its mode shape (Fig. 9) illustrates the alternating process of roll-up to supreme precision. The shear layers enclosing the recirculation region gain curvature by shearing with the external high-speed flow. The curvature is further promoted as the negative base pressure draws in the shear layers. Excessive curvature triggers the roll-up of the shear layers into Strouhal vortices. Thereafter, the rolled-up vortices saturate and break the turbulent sheet before shedding into the near wake. Cogently, Mode 2 describes the phenomenology so accurately that it almost replicates the schematic illustration [51]. On this basis, Mode 2 describes the physics of the Strouhal vortex.

On the other hand, the oscillatory Mode 3 describes the periodic formation and transport of the BG vortices. Fig. 10 illustrates a sequence of coherent structures alongside the shear layers. They are smaller in size and more in number compared to the Strouhal vortices. They are also belted into the near wake by the shear layers, destining to the same locations where the Strouhal vortices roll up for an inevitable superimposition. The region marked in blue is also clear evidence of fluid entrainment. Precisely, it is the engulfment of irrotational fluid into the cavities of time-averaged production [52]. Again, except flipped in orientation, the mode shape matches the quintessence of the schematic illustration and photo footage from Unal and Rockwell [49]. On this note, the Koopman analysis reveals that the BG vortex and the rib-roll substructures appear in not only the cylinder wake but also the prism wake.

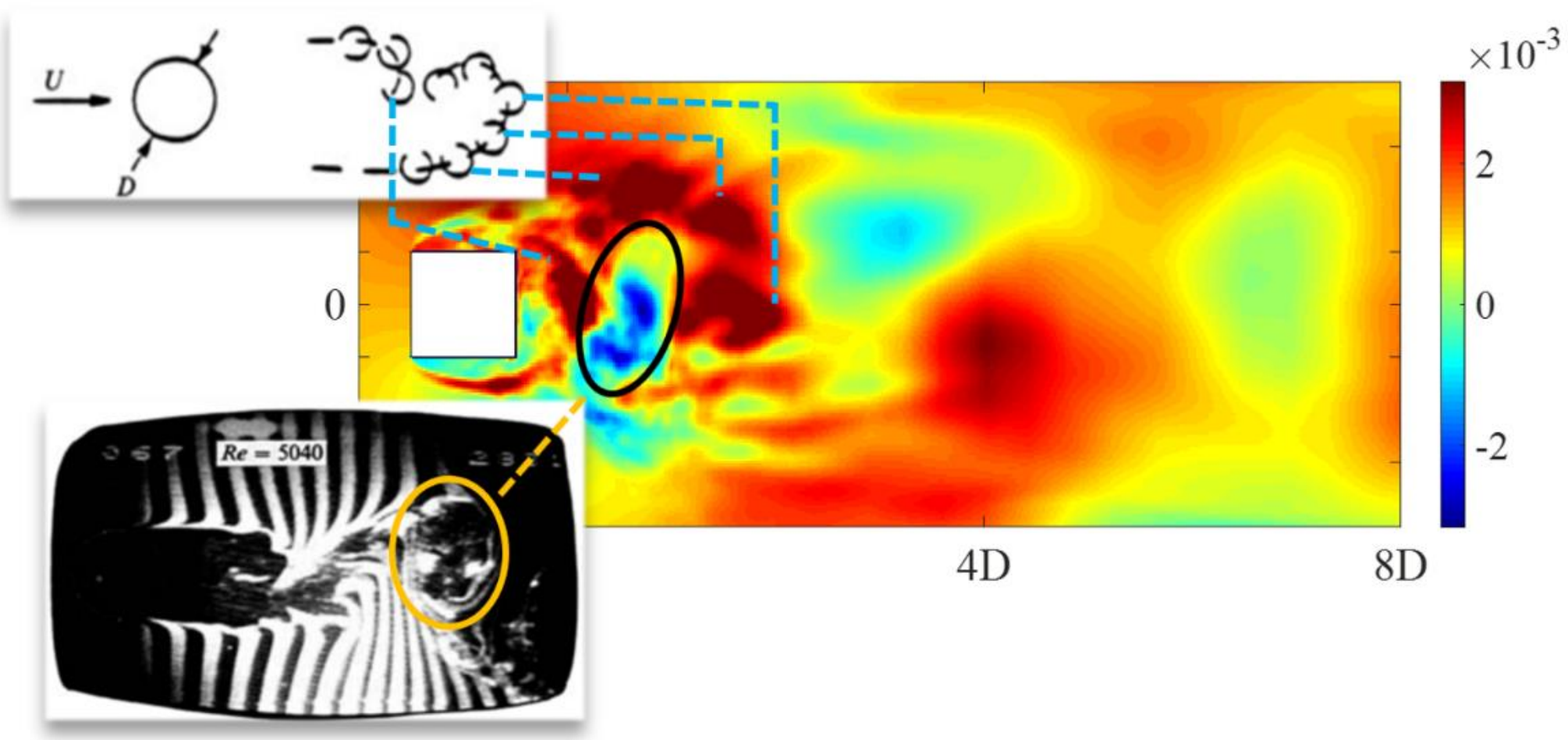

**Figure 10** Mode shape of Mode 3 compared to the schematic illuration and visual footage of the formation of Bloor-Gerrard vortices from Unal and Rockwell [49] ($St_3 = 0.131$)

On the temporal note, Mode 3 decays faster than Mode 2 (Table 1). Unal and Rockwell [49] and Wu *et al*. [43] both reported that the critical Reynolds number for the BG vortices is much higher than that of the Strouhal vortices. This is to say, the mechanism depicted by Mode 3 requires much higher energy to sustain. Therefore, in the absence of continuous excitations, Mode 3 will indeed decay faster. More importantly, to the best of our knowledge, while Mode 2 is ubiquitously reported in the DMD literature [3], Mode 3 is rarely detected. The explanation is that most existing DMD works on the flow over bluff-body configuration consider the low-*Re* number regime, typically $Re_L < 500$. The low $Re_L$ is sufficient to trigger the Strouhal vortices [42] but not the BG vortices at $Re_L > 1900$ [49]. The absence of the BG vortices makes the detection of Mode 3 impossible.

#### *4.2.3 Harmonic Excitation – Modes 4, 5, 7, 8 and 9*

Modes 4, 5, 7, 8, and 9 describe the harmonic excitations of Modes 2 and 3. Specifically, Modes 4, 5, and 9 correspond to the 8$^{th}$, 5$^{th}$, and 2$^{nd}$ harmonics of Mode 2, respectively. Likewise, Modes 7 and 8 correspond to the 4$^{th}$ and 3$^{rd}$ harmonics of Mode 3. The time evolutions (Fig. 11) show all harmonic Modes are oscillatory therefore periodic. The growth rates suggest that the higher the harmonic, the faster the decay, which obeys the energy characteristics of harmonic excitations. Spatially, the mode shapes are also typical of harmonic excitations and agree well with those reported in previous studies [3], [38].

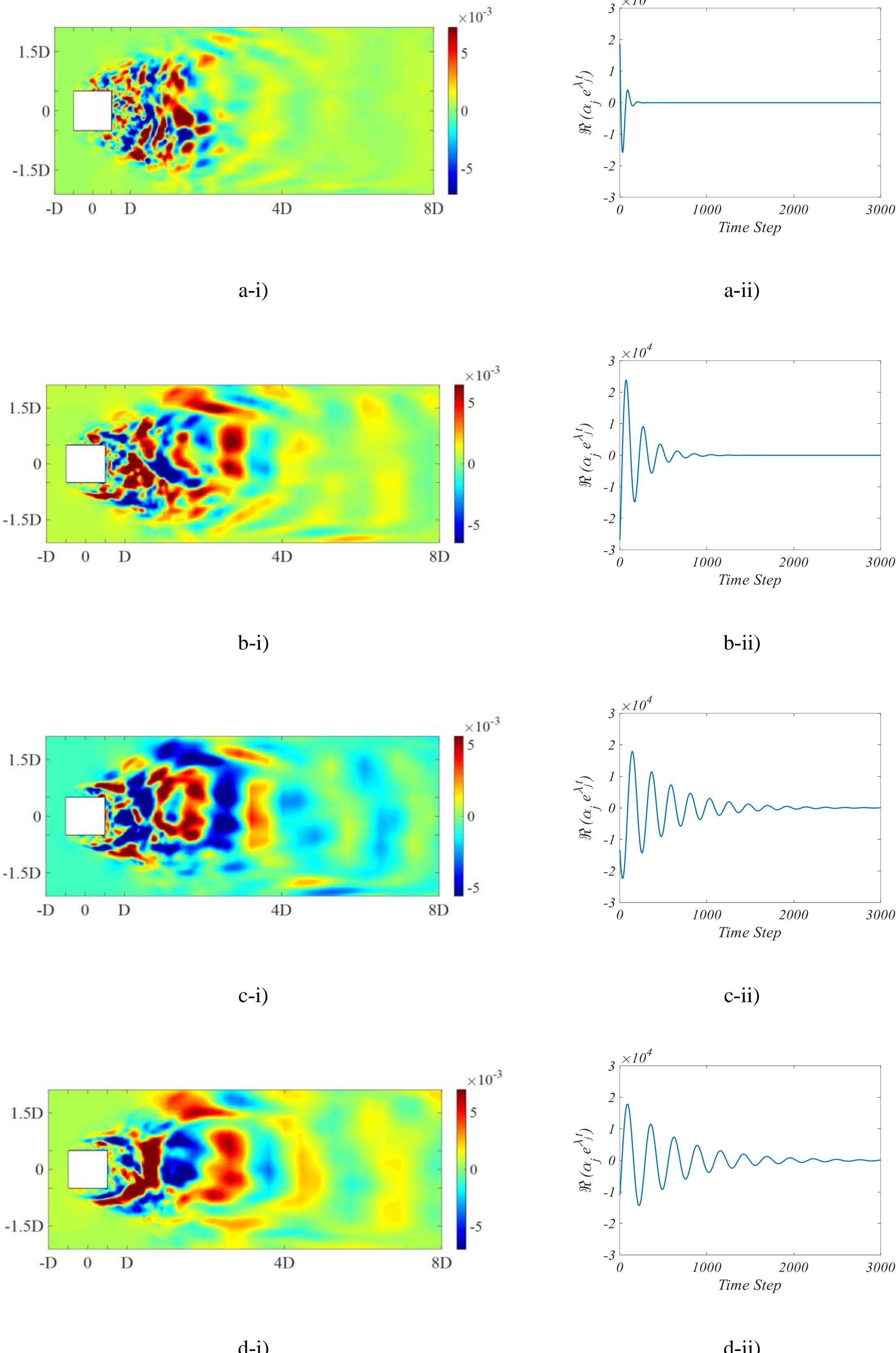

a-i) a-ii)

b-i) b-ii)

c-i) c-ii)

d-i) d-ii)

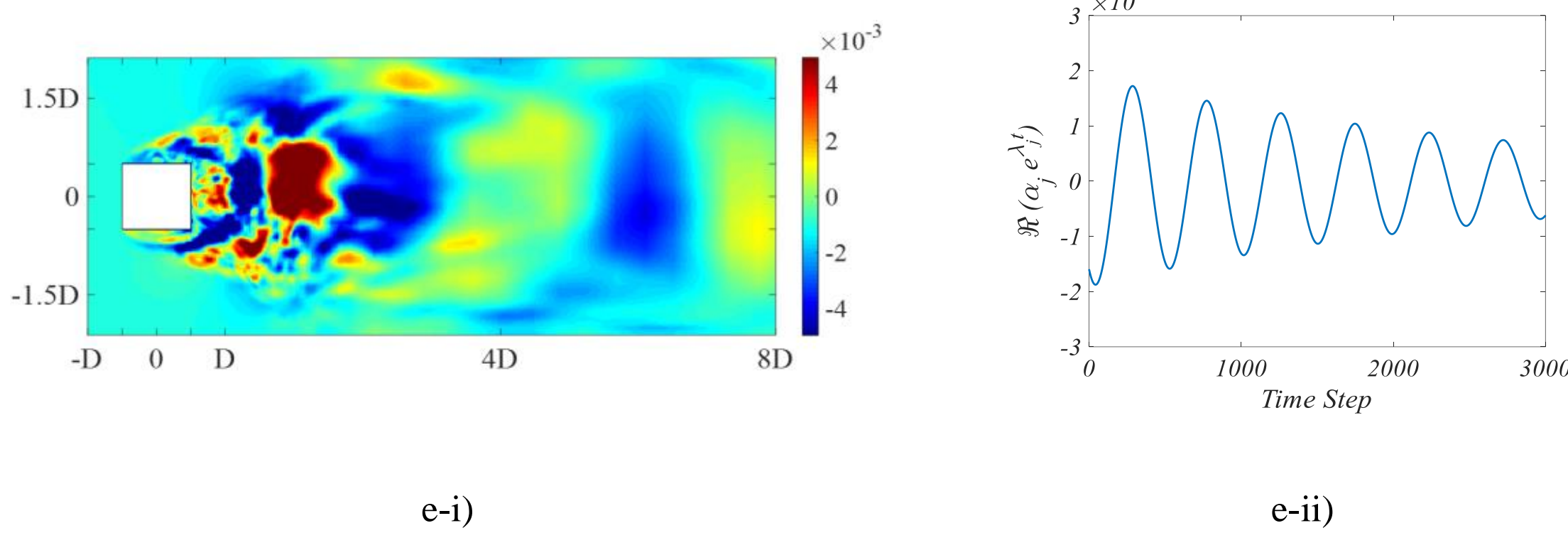

**Figure 11:** Mode shape (i) and time evolution (ii) of Modes **a)** 4 ($St_4$ = 1.067), **b)** 5 ($St_5$ = 0.636), **c)** 7 ($St_7$ = 0.560), **d)** 8 ($St_8$ = 0.466), and **e)** 9 ($St_9$ = 0.256).

We remind the readers of the dubious stability of Mode 4. Intuitively, it is unusual to excite the 8$^{th}$ harmonic intensely that it has the 4$^{th}$ greatest contribution to the overall dynamics. Inspecting its time evolution (Fig. 11a-ii), Mode 4 decays very fast to a negligible amplitude. So, we infer that the 8$^{th}$ harmonic happens to be highly excited in the first few snapshots of the data sequence. The fortuitous choice of the sampled snapshots led to Mode 4's short-lived modal dominance. This abnormality disappeared with an increased sampling sequence, as we will discuss subsequently.

*4.2.4 Subharmonics– Modes 6 and 10*

Modes 6 and 10 have unexpectedly low frequencies, $St_6 = 0.072$ and $St_{10} = 0.051$ respectively, compared to $St_{vs}$. Both Modes are oscillatory therefore describe periodic phenomena (Fig. 12). Incidentally, Muld *et al*. [33] and Cesur *et al*. [34] also observed these low-frequency Modes. The former associated them with shear layer flapping due to an unsteady inlet behaviour. This notion was dismissed by the latter, who discovered a fundamental low-frequency Mode with two associated derived Modes. Cesur *et al*. [34] concluded that the combination of three Modes gave rise to a beating phenomenon, a force amplification, and wake flapping motions. According to their study, the Strouhal number of the fundamental low-frequency Mode $St_{LF}$ can be related to those of the two derived Modes by $St_D \approx St_{vs} \pm St_{LF}$. In view of that, treating Mode 6 as the fundamental low-frequency Mode, $St_D \approx 0.126 \pm 0.072$, the two derived Modes are Mode 10 $St_{10} = 0.051$ and Mode 19 $St_{19} = 0.196$.

The fundamental low-frequency Mode causes strong interactions between the BKVS and the beating phenomenon. To this end, Mode 6 is responsible for the beating phenomenon in instantaneous lift (Fig. 3e) and low-frequency peak in the PSD periodogram (Fig. 4). Cesur *et al*. [34] also discovered the low-frequency activity is responsible for the flapping motions of the shear layers and prism wake. However, despite the valuable observations, the origin of the low-frequency activity remains unexplored.

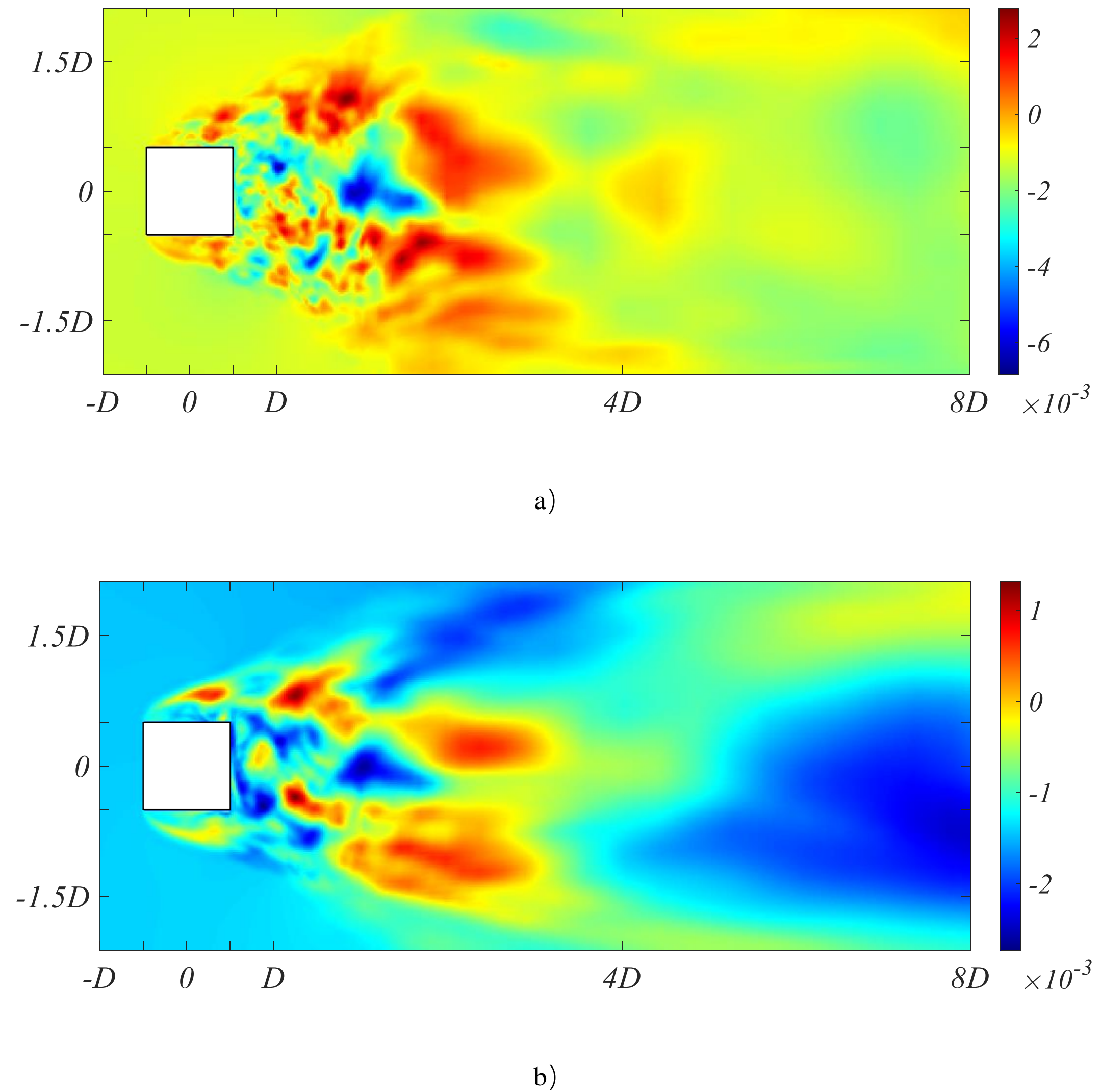

**Figure 12** Mode shape of Modes **a)** 6 ($St_6 = 0.072$) and **b)** 10 ($St_{10} = 0.051$).

The oscillatory Modes 6 and 10 describe the subharmonics of the primary structure. Ducoin *et al.* [53] employed the DMD to study the phenomenology of the SD7003 airfoil wake at $Re=20,000$ and also found the subharmonic peak. They attributed the subharmonic to the intermittent shedding of the laminar separation bubbles (LSB) shedding at 1/2 $St_{vs}$, which also led to a similar beating phenomenon in the time history of lift coefficient. To this end, a boundary layer forms one the airfoil, so the separation bubble is initially laminar, which, of course, is subjected to the ensuing laminar-turbulent transition. However, a boundary layer does not exist for the prism because the sharp leading corners impose forced separation. Therefore, if we attribute the subharmonics to the separation bubbles, then it will be the intermittent shedding of the turbulent separation bubbles (TSB), which has been reported by Kiya and Sasaki [54]. They found the flow unsteadiness generates an accumulation of vorticity inside the TSBs, so small-scale coherent structures formed and coalesced before periodically shedding into the near wake. The intermittent shedding of the TSB is coupled with flapping motions of shear layers, as the initial vorticity accumulation led to a bulking of the TSBs and the ensuing shedding caused a sudden shrinkage. Modes 6 clearly describes the shedding process of the separation bubbles.

Nevertheless, on contrary to association by Cesur *et al*. [34] we dissociate Modes 6 and 10 in their phenomenological implication. The tail-like coherent structures of Mode 10 are, as illustrated by Hussain [44] with a turbulent jet, characteristic of time-averaged production, which mainly consists of shear production with sufficient convection . In a turbulent free-shear flow, the mean velocity gradient and the mean momentum transfer are mostly like-sign, resulting in positive production that generates the tail structures. If we treat the prism wake as two asymmetric wall jets, therefore, two stacked mixing layers, the two tails in Mode 10 are expected. The origin of production traces back to vortex stretching. Its effect can be summarized as the consistent thinning, on statistical average, of fluid elements in the direction perpendicular to the stretching, therefore reducing the radial length scale of the associated vorticity or vortical structures, and ultimately drives the downward cascade into the dissipative scales. Vortex stretching is omnipresent in turbulence.

### *4.3 DMD Sampling and Stability*

We dedicate this final section to a discussion on DMD sampling and stability. To the best of the authors' knowledge, a criterion for DMD sampling was not proposed by the introduction of the DMD [1], nor in the works of the technique's strongest advocates [2], [3], [6] --- there is yet a commonly accepted procedure for DMD sampling. Table 2 summarizes the status of inconsistency. The data sampling of previous work varied from 1 to 270 vortex shedding cycle(s) (*i.e.*, 150 snapshots to 30,000 snapshots) and resolved each oscillation cycle by 17 to 50 time-steps.

**Table 2** Examples of DMD implementations on fluid systems with various sampling range and resolution.

| *Sampling Range [snapshots]* | *Sampling Resolution** | *Configuration* | *Contribution* |
|---|---|---|---|
| 251 | 2 $t_{DNS}$ | Jet flow | Rowley *et al.* [9] |
| 20-30 | u.c. | Prism wake | Schmid *et al.* [1] |
| 500 | 2000 Hz | Flexible membrane wake | |
| 280 | 280 f.p.c. | Cylinder wake | Jardin and Bury [55] |
| 3000 | 20 $t_{DES}$ | Wall-mounted cube | Muld *et al.* [33] |
| 89 cycles | 112 f.p.c. | Cylinder wake | He *et al.* [56] |
| 1000 | 1000 Hz | Cylinder wake | Tissot *et al.* [57] |
| 896 | 40 f.p.c. | Wind turbine | Sarmast *et al.* [58] |
| 1200 | 45 $t_{washout}$ | Pipe flow | Gómez *et al.* [59] |
| 501 | u.c. | Transitional Jet | Roy *et al.* [60] |
| 123 | 30 Hz | Bluff-body wakes | Wan *et al.* [61] |
| 300 | 17 f.p.c. | SD7003 airfoil | Ducoin *et al.* [53] |
| 300 | u.c. | Transonic backward step | Statnikov *et al.* [62] |

*u.c. – unclear; f.p.c – frames per cycle

In this work, we kept the DMD sampling consistent with the LES time-step to ensure the optimal resolution of the system's dynamics. Inevitably, the sampling range is limited to three oscillation cycles considering computational expenses. However, to cast away any doubts on stability, the original dataset (Dataset 1) is supplemented by another dataset (Dataset 2) sampling 18 oscillation cycles. A comparison identifies all the dominant Modes across mean-included datasets (Table 3). The ranking of the Modes varies but is within expectation. The frequency of the TSB shedding tends closer $1/2St_{vs}$, as suggested by [54]. However, considering the vast amount of HPC resources expended on Dataset 2, the margin of improvement is limited for this specific rendering.

**Table 3** Comparison of dominant DMD modes with 3 and 18 sampled oscillation cycles.

| **Dataset 1**: 3 Oscillation cycles | | **Dataset 2**: 18 Oscillation cycles | |
|---|---|---|---|
| ***St*** | ***Remark*** | ***St*** | ***Remark*** |
| 0 | Mean-field | 0 | Mean-field |
| 0.126 | Strouhal vortex | 0.127 | Strouhal vortex |
| 0.131 | BG vortex | 0.121 | BG vortex |
| 1.067 | Harmonic excitation | 0.063 | TSB shedding |
| 0.636 | Harmonic excitation | 2.861 | Harmonic excitation |
| 0.072 | TSB shedding | 0.370 | Harmonic excitation |
| 0.560 | Harmonic excitation | 0.059 | Turbulence production |
| 0.466 | Harmonic excitation | 0.303 | Harmonic excitation |
| 0.256 | Harmonic excitation | 0.927 | Harmonic excitation |
| 0.051 | Turbulence production | 1.494 | Harmonic excitation |

# 5. Concluding Remarks

This work applied the vanilla DMD algorithm to an LES-generated turbulent free-shear flow at *Re=22,000*. Analysis of the ten dominant DMD Modes successfully discloses five prominent excitation mechanisms. Mode 1 describes the mean-field. The Bérnard-Kármán vortex shedding (BKVS) consists of two distinct mechanisms: the Strouhal vortices described by Mode 2 and the Bloor-Gerrard vortices described by Mode 3. The superimposition of the two, with flow entrainment in tandem, culminate into the Kármán vortex street. Modes 4, 5, 7, 8, and 9 depict harmonic excitations. Modes 6 and 10 illustrate the low-frequency shedding of turbulent separation bubbles (TSB) and turbulence production, respectively, which happen concurrently but independently from the BKVS. TSB is also responsible for the beating phenomenon and shear layer flapping motions in the global flow state. Overall, the DMD's descriptions of fluid dynamics prove both morphologically accurate and physically insightful, exhibiting the technique's great potential in dissecting complex fluid systems for enhanced understandings of bluff-body aerodynamics.

## Acknowledgement

We give a special thanks to the IT Office of the Department of Civil and Environmental Engineering at the Hong Kong University of Science and Technology. Its support for installing, testing, and maintaining our high-performance servers is indispensable for the current project.

## Declaration

*To be used for all articles, including articles with biological applications*

## Funding

The work described in this paper was supported by the Research Grants Council of the Hong Kong Special Administrative Region, China (Project No. 16207719).

## Conflict of Interest

The authors declare that they have no conflict of interest.

## Availability of Data and Material

The datasets generated during and/or analysed during the current work are restricted by provisions of the funding source but are available from the corresponding author on reasonable request.

## Code Availability

The custom code used during and/or analysed during the current work are restricted by provisions of the funding source.

## Author Contributions

All authors contributed to the study conception and design. Funding, project management, and supervision were performed by K.T. Tse and Gang Hu. Material preparation, data collection, and formal analysis were performed by Cruz Y. Li and assisted by Lei Zhou. The first draft of the manuscript was written by Cruz Y. Li and all authors commented on previous versions of the manuscript. All authors read and approved the final manuscript.

## Compliance with Ethical Standards

All procedures performed in this work were in accordance with the ethical standards of the institutional and/or national research committee and with the 1964 Helsinki declaration and its later amendments or comparable ethical standards.

## Consent to Participate

Informed consent was obtained from all individual participants included in the study.

## Consent for Publication

Publication consent was obtained from all individual participants included in the study.

# Appendix A – Formulation of the vanilla Dynamic Mode Decomposition

## *A1. Data Pre-Processing*

Two sets of flow field data obtained from either experiments or numerical simulations with a uniform sampling time $\Delta t$, after possible noise elimination or extrapolation, shall be arranged into two matrices in snapshot sequences:

$$\boldsymbol{X_1} = \{\boldsymbol{x_1}, \boldsymbol{x_2}, \boldsymbol{x_3}, \ldots, \boldsymbol{x_{m-1}}\} \tag{A.1.1}$$

$$\boldsymbol{X_2} = \{\boldsymbol{x_2}, \boldsymbol{x_3}, \boldsymbol{x_4}, \ldots, \boldsymbol{x_m}\} \tag{A.1.2}$$

where $\boldsymbol{x_i} \in \mathbb{C}^N$ denotes the $i^{th}$ snapshot of the flow field, $M$ denotes the length of sampled snapshot sequence, and $N$ denotes the number of data entries per snapshot.

$\boldsymbol{X_1}$ and $\boldsymbol{X_2}$ can be connected by a mapping matrix $\boldsymbol{A}$

$$\boldsymbol{X_2} = \boldsymbol{A}\boldsymbol{X_1} \tag{A.1.3}$$

where $\boldsymbol{A}$ is a best-fit linear operator. If a system is linear, $\boldsymbol{A}$ provides an exact description of its dynamics. Conversely, if a system is nonlinear, $\boldsymbol{A}$ then becomes a global, spatiotemporal, and linear approximation of its dynamics.

## *A2. Similarity Expression of DMD*

All steps in the similarity matrix expression aim to approximate the mapping matrix $\boldsymbol{A}$ by a similar matrix. First, the economy-sized Singular Value Decomposition (SVD) shall be performed on $\boldsymbol{X_1}$:

$$\boldsymbol{X_1} = \boldsymbol{U\Sigma V^T} \tag{A.2.1}$$

where $\boldsymbol{U} \in \mathbb{C}^{n\times r}$ contains spatially orthogonal modes $\boldsymbol{u_j}$ on an optimal POD subspace; $\boldsymbol{\Sigma} \in \mathbb{C}^{r\times r}$, a diagonal matrix, contains singular values $\sigma_j$ that describe the modal energy of $\boldsymbol{u_j}$; $\boldsymbol{V} \in \mathbb{C}^{m\times r}$ contains temporally orthogonal modes $\boldsymbol{v_j}$ which pertains to the evolution of $\boldsymbol{u_j}$. $\boldsymbol{u_j}$ and $\boldsymbol{v_j}$ are both orthogonal, such that $\boldsymbol{U^T U} = \boldsymbol{I}$ and $\boldsymbol{V^T V} = \boldsymbol{I}$. The superscript $^{\boldsymbol{T}}$ denotes the conjugate transposition. Note truncation in the SVD is only optional, where $r$ denotes the truncation rank.

Thereafter, a POD-projected similar matrix $\tilde{\boldsymbol{A}} \in \mathbb{C}^{r\times r}$ relates to $\boldsymbol{A}$ by:

$$\boldsymbol{A}=\boldsymbol{U}\tilde{\boldsymbol{A}}\boldsymbol{U}^{\boldsymbol{T}} \tag{A.2.2}$$

Then, minimizing the Frobenius norm of the difference between $\boldsymbol{X_2}$ and $\boldsymbol{AX_1}$ to find the optimal subspace:

$$\underset{\boldsymbol{A}}{minimize}\ \|\boldsymbol{X}_2\text{-}\boldsymbol{AX}_1\|_F^2 \tag{A.2.3}$$

which can be expressed as:

$$\underset{\tilde{\boldsymbol{A}}}{minimize}\ \left\|\boldsymbol{X}_2\text{-}\boldsymbol{U}\tilde{\boldsymbol{A}}\boldsymbol{\Sigma}\boldsymbol{V}^{\boldsymbol{T}}\right\|_{\boldsymbol{F}}^2 \tag{A.2.4}$$

Then, $\tilde{\boldsymbol{A}}$ can be obtained by:

$$\boldsymbol{A} \approx \tilde{\boldsymbol{A}}=\boldsymbol{U}^{\boldsymbol{T}}\boldsymbol{X}_2\boldsymbol{V}\boldsymbol{\Sigma}^{-1} \tag{A.2.5}$$

Now, $\tilde{\boldsymbol{A}}$ can be substituted into Eq. (A.1.3) in lieu of $\boldsymbol{A}$ to describe the system's dynamics.

## *A3. Modal Characterization*

To extract spatial and temporal information from $\tilde{\boldsymbol{A}}$, an eigen decomposition shall be performed:

$$\tilde{\boldsymbol{A}}\boldsymbol{W}=\boldsymbol{W}\boldsymbol{\Lambda} \tag{A.3.1}$$

where $\boldsymbol{W}$ contains the eigenvectors $\boldsymbol{w_j}$, and $\boldsymbol{\Lambda}$ contains the corresponding eigenvalues $\boldsymbol{\lambda_j}$. Thereafter, exact DMD modes**,** which describe the exact eigenvectors of $\boldsymbol{A}$, can be obtained by:

$$\boldsymbol{\Phi}=\boldsymbol{X}_2\boldsymbol{V}\boldsymbol{\Sigma}^{-1}\boldsymbol{W} \tag{A.3.2}$$

where $\boldsymbol{\Phi}$ contains the mode shape $\boldsymbol{\phi_j}$.

Every mode $\phi_j$ corresponds to a physical frequency $\omega_j$ in continuous time:

$$\omega_j = \Im \left\{ log(\lambda_j) \right\} / t^* \qquad (A.3.3)$$

and a modal growth rate $g_j$:

$$g_j = \Re \left\{ log(\lambda_j) \right\} / t^* \qquad (A.3.4)$$